# KiWi: A Scalable Subspace Clustering Algorithm for Gene Expression Analysis

Obi L. Griffith[†], Byron J. Gao[‡], Mikhail Bilenky[†], Yuliya Prychyna[†], Martin Ester[§] and Steven J.M. Jones[†]

[†] BC Cancer Agency, Canada    [‡] Texas State University – San Marcos, USA    [§] Simon Fraser University, Canada

obig@bcgsc.ca, bgao@txstate.edu, mbilenky@bcgsc.ca, y_prychyna@yahoo.ca, ester@cs.sfu.ca, sjones@bcgsc.ca

Subspace clustering has gained increasing popularity in the analysis of gene expression data. Among subspace cluster models, the recently introduced order-preserving sub-matrix (OPSM) has demonstrated high promise. An OPSM, essentially a pattern-based subspace cluster, is a subset of rows and columns in a data matrix for which all the rows induce the same linear ordering of columns. Existing OPSM discovery methods do not scale well to increasingly large expression datasets. In particular, *twig clusters* having few genes and many experiments incur explosive computational costs and are completely pruned off by existing methods. However, it is of particular interest to determine small groups of genes that are tightly coregulated across many conditions. In this paper, we present KiWi, an OPSM subspace clustering algorithm that is scalable to massive datasets, capable of discovering twig clusters and identifying negative as well as positive correlations. We extensively validate KiWi using relevant biological datasets and show that KiWi correctly assigns redundant probes to the same cluster, groups experiments with common clinical annotations, differentiates real promoter sequences from negative control sequences, and shows good association with cis-regulatory motif predictions.

## I. Introduction

Numerous studies have used coexpression of large expression datasets to infer functional associations between genes [1], to identify groups of related genes that are important in specific cancers or represent common tumour progression mechanisms [2], to study evolutionary change [3], for integration with other large-scale datasets [4][5], [6], and for the generation of high-quality biological interaction networks [7][8][9] [10]. A number of studies have also attempted to use coexpression to identify coregulation with the hypothesis that if two or more genes are expressed at the same time and location and at similar levels then they may be regulated by the same transcription factors and regulatory elements. This approach has shown promise particularly in simpler model organisms such as *A. thaliana* and *S. cerevisiae* [11] [12][13] [14] and many groups are currently working on implementing this idea in mammalian systems. However, traditional clustering methods have not worked particularly well on large datasets for this problem. Most methods assign each gene to only one cluster while in reality many genes likely take part in multiple processes. Also, global coexpression is measured across all conditions, whereas, it is probable that most genes are only tightly coregulated under certain conditions or locations.

In recent years, a new field of clustering analysis termed subspace clustering (or biclustering) has gained increasing popularity in the analysis of gene expression data and other biological data [15][16][17][18] [19]. In contrast to traditional clustering methods such as hierarchical clustering, subspace clustering methods do not require expression to be correlated across all conditions for genes to be assigned to the same cluster. This has several advantages for data in which biologically relevant subsets exist (e.g. different tissue types) or where a few noisy experiments might significantly bias the results of the clustering algorithm. This also allows assignment of genes to multiple clusters for different subsets of experimental conditions.

More recently, the order-preserving sub-matrix (OPSM) has been introduced and demonstrated as a biologically meaningful subspace cluster model [15] [20]. An OPSM, essentially a pattern-based subspace cluster, is a subset of rows and columns in a data matrix for which all the rows induce the same linear ordering of columns. In terms of gene expression, an OPSM might represent a group of coregulated genes whose expression levels rise and fall synchronously in response to a series of environmental or cellular stimuli.

A recent report reviewed several existing biclustering methods [17]. They found that in general, biclustering methods outperform global methods such as hierarchical clustering. They also showed that OPSM had the highest proportion of clusters with significant enrichment of one or more Gene Ontology (GO) categories and had good correspondence with known pathways according to their analysis of *A. thaliana* metabolic pathways and *S. cerevisiae* protein-protein interaction networks. However, they state that there are considerable performance differences between the tested methods. Performance is a significant factor as the size of the subspace clustering search space (searching all possible subspaces) is nearly infinite and increases exponentially with the size of the dataset to be analyzed. As costs of expression analysis continue to decrease, the numbers and sizes of expression datasets have grown at an ever-increasing rate. The Gene Expression Omnibus (GEO) currently holds over 70,000 samples for over 100 different organisms [21] and the Stanford Microarray Database (SMD) contains over 10,000 public experiments for over 20 organisms [22]. Furthermore, as array designs continue to improve, it has become possible to include probes for essentially all known genes for many species. The development of exon arrays, alternative splicing arrays, whole-genome tiling arrays, SAGE-type experiments, and high-throughput sequencing technologies increases the size of the problem even further. Thus, with expression data of potentially tens or hundreds of thousands of both rows and columns we need algorithms that can handle not only 'large datasets' but 'massive datasets'.

In our experience, we have found that existing subspace clustering methods do not scale well to these larger expression datasets. We have attempted to analyze large datasets with a Gibbs sampling based biclustering algorithm called gene expression mining server (GEMS) [18]; an adaptive quality-based clustering algorithm [16]; an OPSM-based (OP-cluster) algorithm[20]; and PrefixSpan [23], one of the fastest sequential mining algorithms. In all cases, there were either built in limits (e.g. OP-cluster is limited to 100 columns and 5000 rows) or practical limitations in terms of memory or processor requirements that made it impossible to obtain results for our large datasets (Table 1). In particular, those "twig clusters" defined here as clusters having small size (few genes) and naturally large dimensionality (many experiments) incur explosive computational costs and would be completely pruned off by existing methods. However, it is of particular interest to biologists to determine small groups of genes that are tightly coregulated under many conditions. Some pathways or processes might require only two genes to act in concert. Thus, there is a clear need for subspace clustering methods that can be run on large datasets and detect these twig clusters.

Previously, we introduced a framework that discovers significant OPSM subspace clusters from massive datasets [24]. Here we present an open-source software implementation of this algorithm called KiWi (version 1.0) that is capable of running on a number of different biologically relevant datasets ranging from small to very large in size. We extensively validate the resulting clusters for these datasets and show that KiWi correctly assigns redundant probes to the same cluster, groups experiments with common experimental annotations (such as tissue source), differentiates real promoter sequences from negative control sequences, and shows good association with de novo motif predictions (cisRED). As subspace clustering methods continue to gain popularity over global clustering methods, simple and scalable software will be needed to handle the challenge of ever-increasing dataset sizes facing the biologist. To this end, we provide source code and a working executable for KiWi to the bioinformatics community (http://www.cs.sfu.ca/~bgao/personal/).

## II. RESULTS

### A. KiWi subspace clustering results

For all three datasets analyzed, KiWi was able to run to completion (after parameter optimization) and produce a large number of clusters. The results are summarized in Table 2 and density distributions for cluster size (number of genes) versus pattern length (number of experiments) plotted in Figures 1-3. A large number of clusters (13412 to 212532) was identified for the three datasets, with a range of sizes and pattern lengths. In general, KiWi appears well suited to identifying smaller clusters with long patterns. For the GPL96, expO, and Cooper datasets the average cluster size was 5.11, 3.89, and 6.79 and the average pattern length was 24.04, 42.48, and 6.85 respectively.

### B. Grouping of probes to common gene identifier

Figure 4 shows the results of the 'probe to common gene' analysis. Using the expO dataset (because it is based on the more current Affymetrix platform) we found that of the 23705 clusters identified by KiWi, 1880 (7.93%) contained at least one pair of redundant probes (i.e. different probes corresponding to the same gene) and on average, KiWi clusters contained 0.177 redundant probe pairs per cluster. This was significantly more than the 24.5 (0.10%) clusters with at least one redundant pair (p<0.0001, 10,000 permutations) and the 0.002 average number of redundant probes per cluster (p<0.0001, 10000 permutations) identified in our random simulations.

### C. Experimental annotation analysis

Figure 5 shows the results of the 'experimental annotation analysis'. This used the expO data because unlike most gene expression datasets, the expO data are accompanied by careful and comprehensive experimental annotations. The graph shows a significant tendency by KiWi to group experimental dimensions with common experimental annotation terms such as tissue source, histology, gender, ethnicity, smoking, or alcohol consumption status (p=0.009). When only tissue source terms were considered, a nearly identical graph was observed with a similar level of significance =0.005 (Figure 6).

### D. Negative control analysis

Figure 7 shows the average number of negative control sequences (for the Cooper dataset) in KiWi clusters for different pattern lengths (number of experiments) and cluster size (number of genes/promoters). The random expectation is that negative sequences will be included at a constant rate based on the proportion of total genes/promoters that are negatives. This is what we observe with the randomly generated clusters having a very constant mean fraction of negative controls for all cluster sizes and pattern lengths. Overall, the fraction of negative control sequences included in KiWi clusters was 0.129. This was significantly lower than the mean fraction of 0.134 observed for random simulations (p<0.001, 1000 permutations). The more genes that form a cluster (share a KiWi pattern), the less likely that cluster is to include negative control sequences. Similarly, the longer the pattern (more experimental dimensions) a cluster has, the less likely that cluster is to include a negative control sequence. When negative control sequences were excluded and clustered separately from real sequences we found that significantly more clusters (with greater cluster size and patterns lengths) were produced for the real data than the negative control data (Figure 8).

### E. cisRED analysis

Figure 9 shows that KiWi clusters for the Cooper promoter dataset are more likely to have promoters that contain similar conserved motifs than randomly grouped genes. The overall mean promoter similarity score for KiWi clusters of 0.339 was significantly greater than the score of 0.296 for random (Wilcox test, $p < 2.2e-16$). As the cluster size increases (more genes) the separation from random increases. This is also true for increasing number of experiments (Figure 10).

## III. DISCUSSION

We have extensively assessed and validated the performance of KiWi, a subspace clustering implementation that is scalable to very large expression datasets (10000s of genes and 1000s of experiments) and able to identify smaller (twig) clusters. An advantage of KiWi is that subspace clusters can be identified for a dataset of virtually any size. By experimenting with the settings for *k* and *w*, the user can balance the number and quality of clusters identified against desired runtime. We chose settings that would produce the best results (in terms of numbers of clusters and maximum pattern length) but still run to completion in ~24-48hours on an ordinary PC. Typically subspace clustering methods are evaluated on significantly smaller datasets and with only preliminary biological assessments such as GO analysis. Here, we go significantly further and demonstrate KiWi's ability to identify biologically interesting clusters (at both the gene level and experiment level) from expression datasets that were, until now, inaccessible to subspace clustering methods.

An initial matter for discussion is the idea that KiWi is able to identify the so-called twig clusters rather than being limited to clusters with very large numbers of genes over small number of experiments. If we plot a density graph of the number of genes versus number of experiments for all clusters we see that there is a bias towards smaller clusters. For example, in the expO data 80% of clusters had 5 genes or less and 97% had 10 genes or less (Figure 2) although some larger clusters were found as well (up to 162 genes in the GPL96 dataset, 23 in the expO dataset, and 69 in the Cooper dataset). These clusters, while small in gene number, are in many cases coexpressed across a large number of experiments. For clusters with 5 genes or less, the number of experiments over which coexpression was observed ranged from 10 to 120 with a mean of 42.48. Similar trends were seen for the GPL96 and Cooper data (Table 2; Figure 1-3). We believe this is a novel contribution to the subspace clustering field. Previous studies have tended to focus on the larger clusters by design or necessity. While these large clusters have been shown to be of interest or value, there is no reason to expect that all or even most biological processes or disease mechanisms would involve tight co-regulation of large groups of genes. In fact, we expect that many important processes will involve relatively small numbers of genes. Indeed, the biological assessments discussed below showed that many of these small clusters are of biological interest.

Gene Ontology (GO) and oPossum analysis results were previously reported [24] using a large Affymetrix dataset (referred to here as GPL96). We showed that KiWi can group genes with over-representation of GO biological processes and oPossum transcription factor binding sites. In another previous study [25] we found that for the GPL96 data, approximately 33% of coexpressed gene pairs with a global Pearson correlation (r) of at least 0.80 shared a common GO biological process term. This was the minimum r value for which we saw clear separation from random performance. Other studies have recommended r=0.84 as a good cutoff for reliable global coexpression [26]. However, even with the less stringent cutoff of r=0.8, only 9701 gene pairs actually attained this level of global coexpression. We hypothesized that subspace clustering might provide a useful alternative or complementary method for identifying biologically relevant coexpression relationships. With KiWi we were able for the first time to analyze our large GPL96 dataset (updated with a further 973 experiments since our previous publication) for subspace coexpression. The entire set of 13412 KiWi clusters (representing 393352 coexpressed gene pairs) had similar levels of GO performance with 23% of clusters having at least one significant GO term (FDR<0.1) and good separation from random expectation. This is perhaps expected given that the vast majority of KiWi clusters have very high correlation of expression between genes across their subset of experiments. For example, in the GPL96 dataset, 90% of KiWi clusters had an average r > 0.95 (Figure 11) whereas only 44 of the gene pairs identified by global methods had r values this high. This demonstrates that whereas few gene pairs or clusters are highly correlated across all conditions/experiments, most genes are highly correlated with one or more other genes across some subset of the conditions.

Using another large Affymetrix dataset (expO) we show that KiWi is also very good at grouping probes for the same gene. In fact, a significant proportion of the total clusters contain 'redundant' probes. In itself, this is not a surprising result, but is an important positive control that shows KiWi correctly identifies logically related genes. However, this also argues for removing or averaging of redundant probes before clustering to avoid wasted computation time when gene clusters (as opposed to probe clusters) are the desired end-product.

The expO dataset was also useful for its evaluation of experiment-level clustering by KiWi. Part of the promise of biclustering methods is that they will identify not only coexpressed genes but also the subset of experimental tissues or conditions under which they are coexpressed. Such biclusters could be of particular value for identifying tissue- or stage-specific coregulation. But, they also allow identification of coregulation for previously unconsidered sample groups. For example, we were able to identify gene clusters specific to gender, smoking and alcohol consumption status. The Expression Project for Oncology and International Genomics Consortium should be applauded for not only making their raw expression data (CEL files) available in GEO but also for providing these detailed and standardized clinical annotations. Almost none of the other datasets in GEO have done so. Such datasets make possible, for the first time, true two-dimensional evaluation of biclustering methods on clinically relevant expression data.

In the Cooper promoter dataset, expression levels were measured by reporter gene assay for a large number of promoter sequences across a set of 16 different cell lines. Thus, the data comes in the normal format of a gene/promoter versus experiment/condition matrix but expression is measured by reporter gene activity (luciferase levels) instead of hybridization intensity (as on a microarray). The Cooper promoter dataset is a relatively small dataset and was chosen for reasons other than its size. However, in subspace clustering problems, even a seemingly small dataset contains a very large set of possible subspaces. For example, for the Cooper dataset with the parameters we chose, there are $3.3 \times 10^{224}$ possible subspace clusters. What makes the Cooper dataset particularly useful is the presence of a large number of negative control sequences (random DNA sequence). We hypothesized that KiWi would be biased against inclusion of these negative control sequences and this is indeed what we observed. Also, both the pattern length and number of genes seem to be strong predictors of how reliable a pattern is. We can in principle use this information to define rules for the minimum length and cluster size in combination needed for a 'reliable' pattern. Visually, a cluster with pattern length of 10+ looks reliable with 3 or more genes, a cluster with pattern length of 9 looks reliable with 4 or more genes, and so on.

Comparing the Cooper promoter KiWi clusters to a cisRED analysis of the Cooper promoter sequences also showed a tendency for coexpressed genes to share common regulatory motifs. This result suggests that KiWi coexpression analysis may be useful in filtering de novo motif predictions and/or selecting coexpressed genes as input for motif discovery. As in the negative control analysis, we found that both the cluster size and number of dimensions were predictors of promoter similarity. These findings confirm the intuitive idea that OPSM patterns are more likely to be real if they are shared by more genes and/or across more experiments. This implies that smaller (twig) clusters will need longer patterns (more experiments) in order to have the same level of confidence as larger clusters. A useful future development of KiWi would be the development of a score or p-value by which clusters could be automatically ranked that takes both the numbers of genes and experiments into consideration.

## IV. CONCLUSION

By design, KiWi is capable of identifying both negative and positive correlations of expression, twig clusters (as small as two genes), and genes that appear in multiple clusters. We have demonstrated that these clusters correctly group related probe sequences, avoid 'contamination' by negative controls and tend to share common biological processes (GO) and common regulatory sequences as defined by both motif-scanning methods (oPossum) and de novo motif prediction methods (cisRED). Finally, over-representation of experimental annotation terms gives hope that tissue- or condition-specific clusters can be defined. These features suggest that KiWi should be useful for a wide range of biological applications and may be of particular use in the identification of novel groups of coregulated genes. To facilitate these applications, we provide all datasets, source code, a software tutorial and a working executable (for Windows® operating systems) to the bioinformatics research community (http://www.bcgsc.ca/platform/bioinfo/ge/kiwi and http://www.cs.sfu.ca/~bgao/personal/).

## V. METHODS

### A. Algorithm

The KiWi algorithm takes as input a standard gene expression data matrix with genes as rows, experiments as columns, and some measure of expression level for each data point. By sorting the gene (row) vectors and replacing the entries with their corresponding experiment (column) labels, the data matrix can be transformed into a sequence database, and OPSM mining can be reduced to a special case of the sequential pattern mining problem [27]. An OPSM subspace cluster is uniquely specified by this sequential pattern and its supporting sequences. In other words, we are looking for a set of genes (the supporting sequences) which have the same linear order of expression values for some subset of experiments (the pattern). Instead of finding a complete set of patterns that are beyond some minimum number of genes, KiWi targets the longest patterns for any fixed number of genes (i.e. the subspace cluster with the most experiments showing the same pattern of expression).

KiWi exploits two parameters $k$ and $w$ to provide a biased testing on a bounded number of candidates, substantially reducing the search space and problem scale. In particular, KiWi performs a level-wise search, where shorter patterns gradually grow into longer patterns level by level. Based on the observation that more frequent sub-patterns are likely to grow into more frequent super-patterns, we keep the top $k$ patterns at each level with the greatest number of supporting genes. These patterns are used to generate candidates for the next level. Based on the observation that a long pattern segments its supporting sequences into small sections, in counting the number of supporting sequences, we only consider a region of width $w$. In other words, only if the new element of a candidate appears in the next $w$ positions of a sequence do we consider the candidate to be supported by the sequence. Other techniques employed by KiWi, such as the choice of ranking statistics, memory management, pattern extension and redundancy removal, are discussed elsewhere in detail [24]. The average case runtime of KiWi is $O(kwn)$, linear in $n$, the number of rows (genes) [24].

KiWi is also the first subspace clustering method to identify anti-correlation as well as correlation. Anti-correlation of expression is interesting because it can also imply common process/pathway membership or negative regulation [28]. Anti-correlated genes might also represent members of opposing pathways (when one is active the other is repressed) [29] or cases where expression of one gene represses the expression of other genes (negative regulators). Anti-correlation in the context of subspace clustering can be captured by the so-called generalized order-preserving sub-matrix (GOPSM) [24], where all the genes in a GOPSM induce the same or opposite linear ordering of experiments. KiWi mines GOPSM subspace clusters by searching the sequence database forward and backward simultaneously. KiWi marks any cluster that contains one or more anti-correlated genes. If an 'anti-correlated cluster' contains more than two genes it will by definition contain both positively and negatively correlated pairs of genes.

### B. Datasets

Several datasets and methods were utilized to test for biological coherence of gene clusters predicted by the KiWi clustering algorithm. Expression datasets utilized include: (1) GPL96 - a set of 1640 Affymetrix (HG-U133A) experiments from the Gene Expression Omnibus (GEO, GPL96) [21] covering a broad range of experimental conditions; (2) expO - a set of 1026 Affymetrix (HG-U133 Plus 2.0) experiments from 123 different cancer tissue types from the expO (Expression

Project for Oncology) project (GEO, GSE2109); (3) Cooper promoters - a high-throughput promoter dataset reported by Cooper et al (2006) consisting of 16 cell lines for which the expression of 632 promoter sequences (plus 98 negative control sequences) were assayed by reporter gene assay [30]. These datasets are summarized in Table 1.

*C. Dataset processing*

The Affymetrix HG-U133A probes were normalized and mapped to gene/protein identifiers as previously reported [25]. The expO data were normalized from CEL files using the Bioconductor 'just.gcrma' function in the 'gcrma' library (version 2.4.1). Probes were mapped to Uniprot and Ensembl identifiers using the Bioconductor 'biomart' package [31]. The Cooper promoter data were used as provided. Tab-delimited data matrices for each dataset were loaded into the KiWi software (v. 1.0) and subspace clusters identified using the parameters outlined in Table 3. Parameters were chosen by experimentation to identify values of $k$ and $w$ that would produce the largest number of clusters and longest patterns but still run to completion within 24 to 48 hours.

*D. Biological evaluation*

**Grouping of probes to common gene identifier.** Affymetrix gene expression chips (such as HG-U133 Plus 2.0, used for expO) contain numerous probe sets derived from the same gene either as redundant probe sets or probe sets for different transcripts of the same gene. It is expected that such probe sets will display correlated expression given that they measure the same or related transcripts. Therefore, we expect that probe sets mapped to a common gene identifier will be grouped together in the same subspace cluster more often than expected by chance. This represents a kind of positive control experiment. The number of probe pairs mapped to the same gene was determined for all clusters. Significance was assessed by random permutation analysis. That is, 10000 sets of clusters were randomly generated (with the same sizes and dimensions as produced by KiWi). The mean number of redundant probe pairs for all clusters was then compared between KiWi and the distribution of random results.

**Experimental annotation analysis**. KiWi subspace clusters consist of a set of 2 or more genes found to have correlated expression patterns (specifically an OPSM) for some subset of the available experiments. Most validation methods look for biologically consistent grouping of genes. To determine if the experiments are also grouped in a meaningful way, an over-representation analysis was applied to experimental annotations. The expO dataset was chosen as this dataset is accompanied by carefully annotated clinical details. For example, each experiment is annotated as one of 123 different tumor tissue types. Clusters with a minimum of two genes and 50 experiments were selected for analysis. Statistically over-represented experimental annotation terms were defined using Fisher Exact statistics and corrected for multiple testing by a Benjamini and Hochberg (BH) correction with the Bioconductor 'multtest' package. Overall significance of the KiWi clusters was determined by comparing to randomly generated clusters using a Kolmogorov-Smirnov test.

**Negative control analysis**. In the Cooper promoter dataset, the authors measured expression levels by reporter gene assay for a large number of promoter sequences across a set of 16 different cell lines. They also included a large number of negative control sequences (random DNA sequence). Unlike the real promoter sequences, we do not expect these sequences to drive gene expression in any meaningful way. They may produce some low level 'noisy' expression values but nothing coordinated. Any clusters formed from such data most likely represent random patterns as opposed to meaningful patterns. Therefore, we can hypothesize that if Kiwi is detecting 'true' or 'real' clusters they should be biased towards positive sequences and against negative sequences. To this end, we ran KiWi on expression data for all sequences (positives and negatives) and determined the fractions of KiWi clusters 'contaminated' by negative control sequences. Significance was determined using a random permutation approach as described above. We also clustered positive and negative sequences separately and compared the numbers, sizes (number of genes) and pattern lengths (number of experiments) of clusters for each dataset.

**cisRED analysis**. For all Cooper promoter sequences (excluding negative control sequences) the cisRED pipeline was used to predict putative regulatory motifs as described previously [32]. Briefly, cisRED uses multiple discovery methods applied to sequence sets that include up to 42 orthologous sequence regions from vertebrates (mostly mammals). Motif significance is estimated by applying discovery and post-processing methods to randomized sequence sets that are adaptively derived from target sequence sets. Motifs are then annotated based on their similarity to known transcription factor binding site (TFBS) models (using TRANSFAC 9.3). Motifs with p-values below a threshold (discovery p-value < 0.001 and annotation p-value < 0.0005) are retained, groups of similar motifs identified and co-occurring motif patterns defined. Any set of two or more genes with a co-occurring motif pattern is hypothesized to be co-regulated by one or more transcription factors. We can further hypothesize that these putatively co-regulated gene groups are more likely to belong to a cluster of coexpressed genes (as defined by KiWi) than randomly formed groups of genes. To test this, a promoter similarity score was defined. For every KiWi cluster, for each pair of genes in the cluster, we calculate the number of cisRED motifs annotated with the same TFBS model (counting repeated annotations only once; Figure 12). More formally, we define $L_i = \{l_{i,}^1, l_{i,}^2, ..., l_i^n\}$ to be a set of annotation labels for $n$ conserved motifs predicted in the promoter of a gene "$i$". Then we define a promoter similarity score (S) as the average number of common annotation labels for every pair ($i, j$) of genes from a KiWi cluster of size N as follows:

$$S = 2 \frac{\sum_{i,j} |L_i \cap L_j|}{N(N-1)}$$

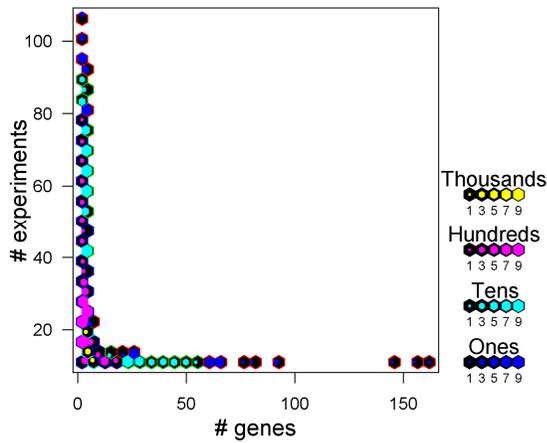

Figure 1. KiWi results for the GPL96 dataset.

The density plot shows the cluster size (number of genes) versus pattern length (number of experiments) for all clusters. The density plot was produced using the Bioconductor 'hexbin' library (version 2.3.0).

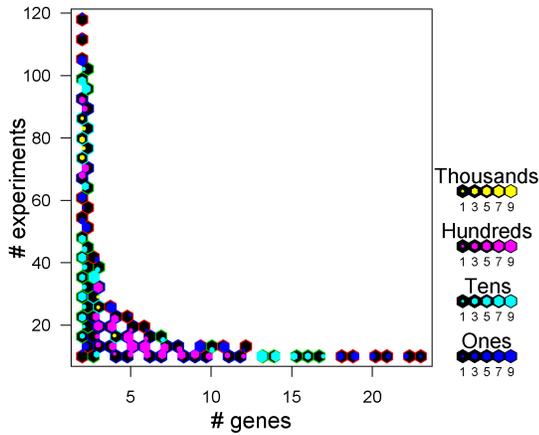

Figure 2. KiWi results for the expO dataset.

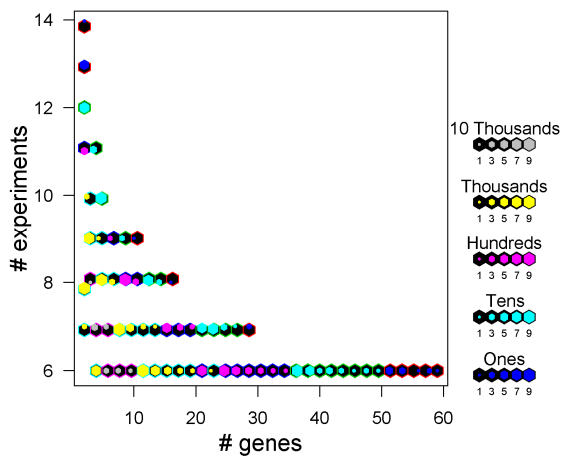

Figure 3. KiWi results for the Cooper dataset.

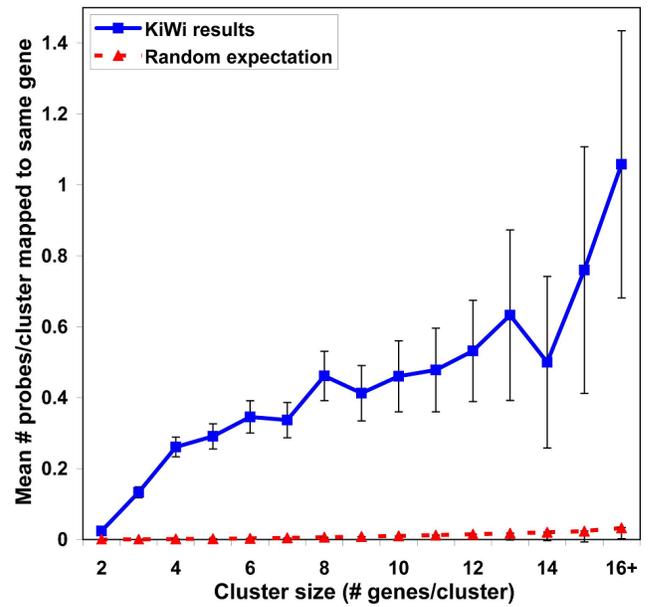

Figure 4. Probe to common gene analysis.

The mean number of probe pairs in a cluster that are mapped to the same gene (redundant probes) is shown for each cluster size (number of genes per cluster). Error bars indicate 95% confidence limits.

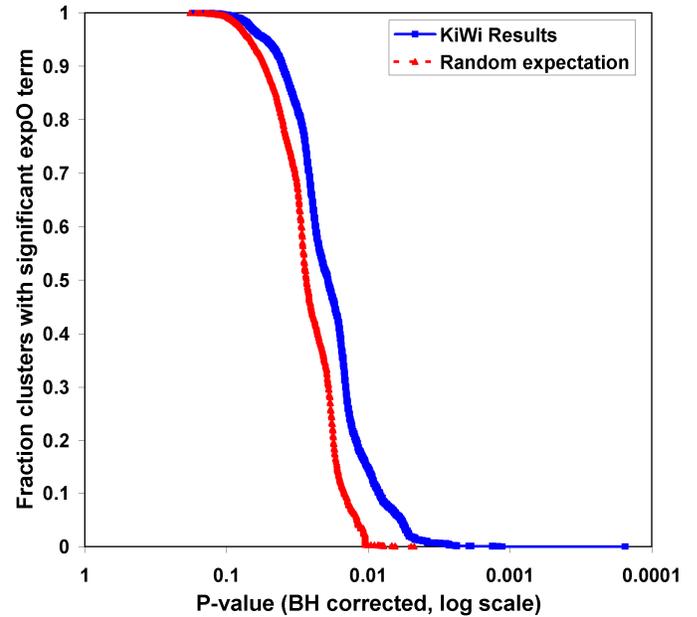

Figure 5. Experimental annotation analysis.

The fraction of clusters with at least one significantly over-represented experimental annotation term at each level of significance is shown. Significance was determined by Fisher Exact test. P-values were corrected by the Benjamini and Hochberg method and are displayed on a reverse log scale.

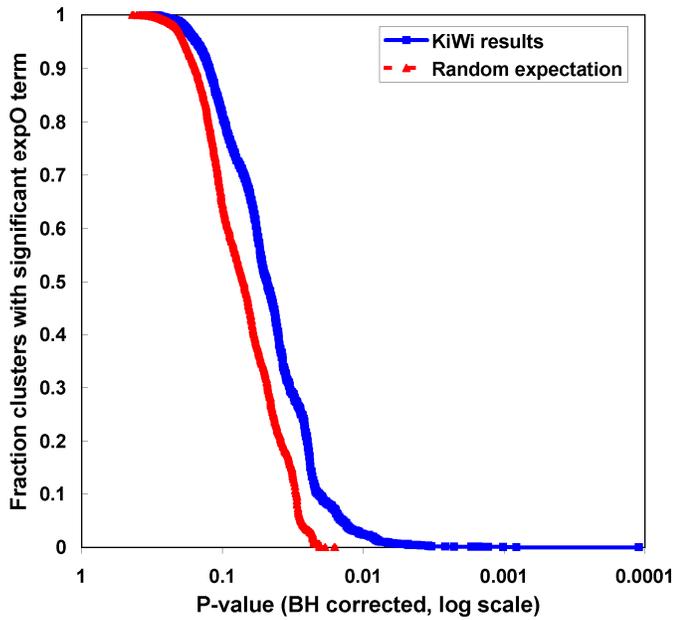

Figure 6. Experimental annotation analysis excluding all annotation terms except tissue source.

The fraction of clusters with at least one significantly over-represented tissue source term at each level of significance is shown. Significance was determined by Fisher Exact test. P-values were corrected by the Benjamini and Hochberg method. Kolmogorov-Smirnov test showed a significant difference from random (p=0.005).

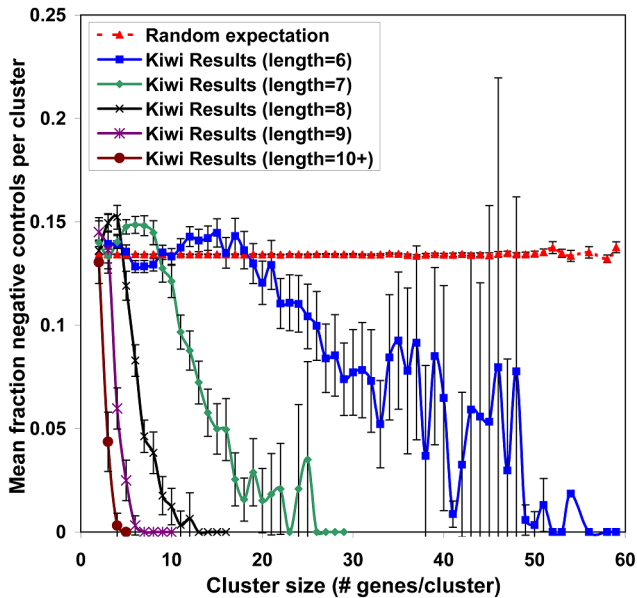

Figure 7. Negative control analysis.

The mean fraction of negative control sequences included in each cluster for each cluster size is shown. Results are broken down by pattern length except for the random results for which all cluster sizes and pattern lengths showed constant contamination by negative controls. Error bars indicate 95% confidence limits.

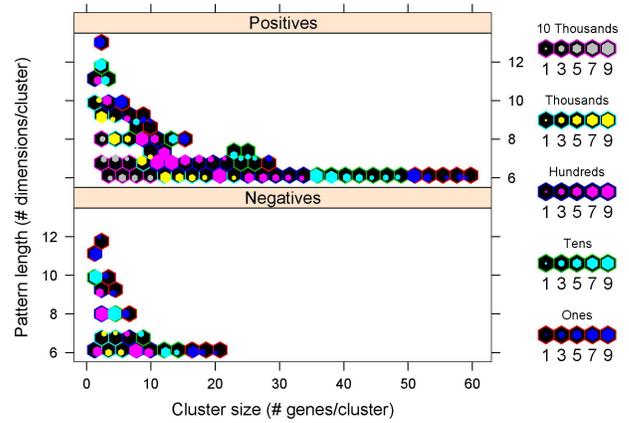

Figure 8. KiWi results for Cooper dataset with negative control sequences (Negatives) and real promoter sequences (Positives) clustered separately.

The two datasets were submitted to KiWi with identical parameters (k=100000; w=16; min_row=2, min_col=6). The density plot shows that the positive data inherently produces more clusters with longer patterns (more experiments) and greater cluster size (more genes).

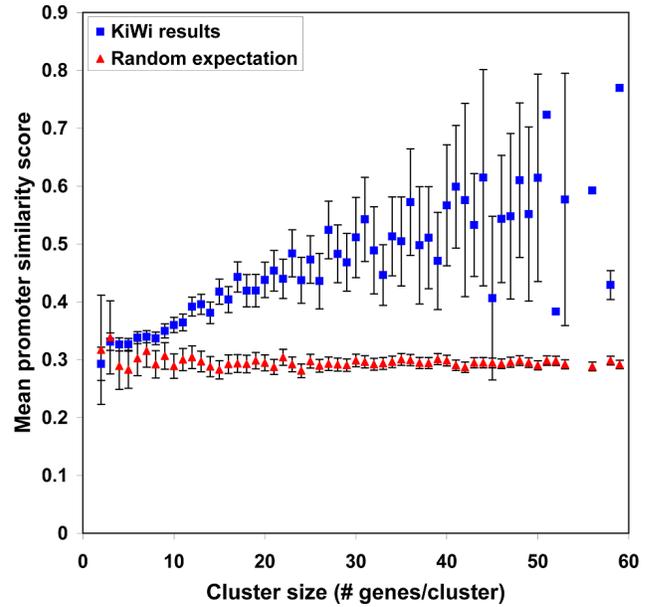

Figure 9. cisRED analysis.

The mean promoter similarity score (see methods) for each cluster size is shown. Random expectation is based on 2000 randomly generated clusters for each cluster size. Error bars indicate 95% confidence limits.

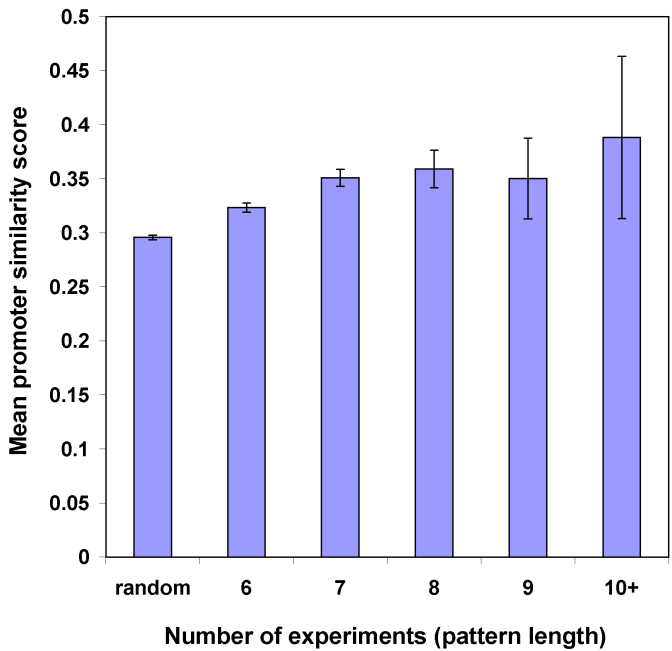

Figure 10. cisRED analysis comparing different pattern lengths.

The mean promoter similarity score for each pattern length is shown. As with cluster size (Fig. 9) the promoter similarity score increases with greater pattern length (number of experiments). Error bars indicate 95% confidence limits.

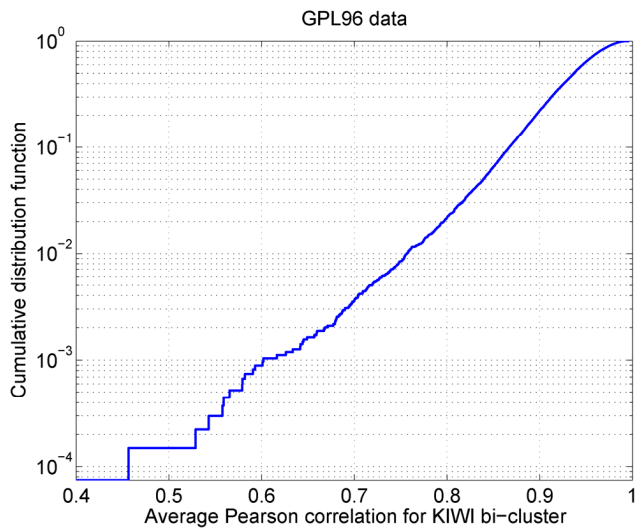

Figure 11. Distribution of mean Pearson correlations for all KiWi clusters for GPL96 data.

This figure shows that the vast majority of subspace clusters (bi-clusters) have very high average pairwise correlations with 90% of clusters having an average r > 0.95.

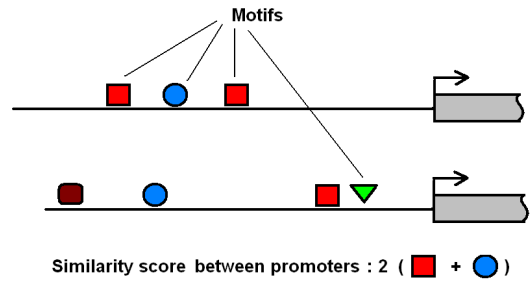

Figure 12. Diagrammatic explanation of "promoter similarity score".

For any pair of genes, the upstream region is compared for overlap of annotated TFBS motifs (Sp1, AP-2, etc). These are represented as colored shapes in the diagram above. Each common motif is counted once. The promoters above share two motifs (red square and blue circle). Therefore, the score for this gene pair is 2. Then, the overall promoter similarity score (S) for a cluster of genes is calculated as the sum of the pairwise scores divided by the number of pairs.

TABLE I. DATASETS USED IN KIWI ASSESSMENT

| Dataset | # of rows | # of columns |
|---|---|---|
| GPL96 | 12332 | 1640 |
| expO | 20113 | 1026 |
| Cooper promoters | 730 | 16 |

TABLE II. KIWI RESULTS

| Dataset | # clusters found | Mean genes/cluster (range) | Mean exps/cluster (range) |
|---|---|---|---|
| GPL96 | 13412 | 5.11 (2 to 162) | 24.04 (11 to 108) |
| expO | 23555 | 3.89 (2 to 23) | 42.48 (10 to 120) |
| Cooper | 212532 | 6.79 (2 to 59) | 6.85 (6 to 14) |

TABLE III. PARAMETERS USED IN KIWI ASSESSMENT

| Dataset | k | w | Min. genes | Min. exps |
|---|---|---|---|---|
| GPL96 | 30000 | 45 | 2 | 10 |
| expO | 100000 | 18 | 2 | 10 |
| Cooper | 100000 | 16 | 2 | 6 |